\documentclass[twocolumn,showpacs,amsmath,amssymb,prb]{revtex4}

\usepackage{graphicx}
\usepackage[usenames]{color}

\newcommand{\up}{\uparrow}
\newcommand{\dn}{\downarrow}
\newcommand{\vareps}{\varepsilon}
\newcommand{\ket}[1]{\left|#1\right>}
\newcommand{\bra}[1]{\left<#1\right|}
\newcommand{\dt}[1]{\frac{\text{d}#1}{\text{d}t}}

\newcommand{\mw}[1]{\left\langle #1 \right\rangle}
\newcommand{\tr}[1]{\mathrm{tr}\left[#1\right]}

\begin{document}

\title{Current fluctuations in noncollinear single-electron spin-valve transistors}

\author{Stephan Lindebaum}
\affiliation{Theoretische Physik, Universit\"at Duisburg-Essen and CeNIDE, 47048 Duisburg}

\author{J\"urgen K\"onig}
\affiliation{Theoretische Physik, Universit\"at Duisburg-Essen and CeNIDE, 47048 Duisburg}

\date{\today}

\begin{abstract}
We present a theoretical framework to analyze fluctuations of the electric current through a noncollinear single-electron spin-valve transistor in the limit of weak tunnel coupling. The system under consideration consists of two tunnel junctions that connect a small, nonmagnetic metallic island to two ferromagnetic leads with noncollinear magnetization. We study the current noise spectrum as a function of bias voltage, frequency, and the relative angle between the leads' magnetization directions and find that both the zero- and the finite-frequency current noise are strongly affected by charging energy and spin accumulation in the island.
\end{abstract}

\maketitle

\section{Introduction} \label{sec:introduction}
The continuous trend of miniaturization in electronics after the realization of the first transistor\cite{bardeen:1948} lead to device dimensions that, nowadays, approach the nanometer scale, at which Coulomb-interaction effects and quantum mechanics become important.
A paradigmatic system for the emergence of such phenomena is the single-electron transistor (SET), in which the continuous movement of charge carriers is replaced by a discrete charging and discharging of a small central electrode.\cite{likharev:1987,goldhaber:1997} 
Electrodes and central island of the SETs maybe composed of different materials, involving normal metals, superconductors, and/or ferromagnets. 
The use of ferromagnetic components give rise to spintronic effects such as the tunnel-magneto resistance (TMR) that can be used in information technology.\cite{datta:1990,wolf:2001,greg:2002,zutic:2004} 
Therefore, SETs based on ferromagnetic materials have been extensively studied experimentally\cite{ootuka:1996,ono:1997,shimada:1998,ono:1998,brueckl:1998,takemura:2001,jedema:2002,shimada:2003,matsuda:2003,niizeki:2004,bernand:2006,wunderlich:2006,seneor:2007,liu:2007,bernand:2009,yakushiji:2001,deshmukh:2002,zhang:2005,sahoo:2005,merchant:2008,hofstetter:2010,holm:2008,darau:2009} and theoretically.\cite{takahashi:1998,barnas:1998a,barnas:1998b,majumdar:1998,korotkov:1999,brataas:1999a,brataas:1999b,barnas:2000,brataas:2001,martinek:2002,weymann:2003,ernult:2007,barnas:2008,brataas:2000,huertas:2000,braig:2005,wetzels:2005,wetzels:2006,linder:2007,lindebaum:2011,bulka:2000,koenig:2003,martinek:2003a,weymann:2005,fransson:2005,weymann:2006,simon:2007,splettstoesser:2008,sothmann:2010,sothmann:2010a,baumgaertel:2010}
Since the signal-to-noise ratio is crucial for possible applications it is important to investigate current fluctuations.
Furthermore, the current noise can reveal additional information about the transport processes that is not contained in the mean current. 
The signal-to-noise ratio is represented by the Fano factor $F=S^{II}/(2eI)$, where $S^{II}$ (properly defined below) is the noise of the current $I$ and $e$ the elementary charge. 
Especially in the context of mesoscopic devices (such as SETs) current fluctuations attracted much interest over the years, for a review see Ref.~\onlinecite{blanter:2000}.

It has been shown that interaction effects can destroy the typical fermionic suppression of the classical Poisson noise. Super-Poissonian Fano factors have been observed in devices containing electron reservoirs that are coupled to quantum wells,\cite{blantner:1999} quantum dots,\cite{sukhorukov:2001,kiesslich:2003,cottet:2004,belzig:2005a,braun:2006,nguyen:2006,bulka:2008,lindebaum:2009,michalek:2009,wang:2011} single molecules,\cite{thielmann:2005,elste:2006,contreras:2010} carbon nanotubes,\cite{thielmann:2005,weymann:2008} single-barrier semiconductor heterostructure,\cite{reklaitis:2000} and quantum rings.\cite{cavaliere:2005}
 
\begin{figure}[b]
		\includegraphics[width=.9\columnwidth]{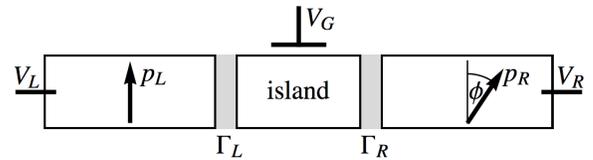}
		\caption{(Color online) A metallic island is tunnel coupled to two adjacent ferromagnetic leads with noncollinear magnetization directions enclosing an arbitrary angle $\phi$. This system is called single-electron spin-valve transistor.}
		\label{fig:system}
\end{figure}

In the present work we study current fluctuations of a single-electron transistor composed of a central metallic island that is weakly tunnel coupled to two ferromagnetic leads, whose magnetization directions enclose an arbitrary angle $\phi$, see Fig.~\ref{fig:system}.
This so called {\it noncollinear single-electron spin-valve transistor} shows typical single-electron and spintronic phenomena, such as Coulomb oscillations, Coulomb blockade, TMR, and spin accumulation. Furthermore, the system exhibits an interaction-induced exchange field that exists between the central region and the leads.\cite{wetzels:2005,wetzels:2006,lindebaum:2011} This fictitious field is evoked by virtual-tunneling processes between the interacting central region and the polarized leads and results in a precession of the accumulated island spin. The existence of such an exchange field in mesoscopic conductors was firstly theoretically described\cite{braun:2004} and experimentally confirmed\cite{pasupathy:2004,hamaya:2007,hauptmann:2008} in the context of quantum dots coupled to ferromagnetic reservoirs.

In literature, there are several publications dealing with current fluctuations in special limits of the single-electron spin-valve transistor. The complexity of the required theoretical description drastically simplifies if one considers unpolarized leads ($p=0$) only. In this limit (two normal leads coupled to a central normal region with a continuous level spectrum, NNN) zero-frequency\cite{hershfield:1993,bagrets:2003,braggio:2006} as well as finite-frequency\cite{korotkov:1994,hanke:1993,hanke:1994,johansson:2002,kaeck:2003} current fluctuations have been addressed theoretically. Since in experiments the shot noise is often superimposed by other sources of noise, e.g. the noise of the used amplifiers or the $1/f$ noise that occurs due to defects in or near the junctions, there are just a few experimental works studying the current fluctuations of the NNN system. Most of them have to restrict their noise measurements to the regime of large bias voltages.\cite{aassime:2001,aassime:2001a,roschier:2004} However, recently Kafanov and Delsing measured the noise of the NNN system over a wide voltage range.\cite{kafanov:2009} Most of the published works considering current fluctuations of single-electron spin-valve transistors, i.e., SETs involving ferromagnets, restrict the investigations to collinear setups ($p\neq 0$, $\phi\in\{0,\pi\}$).\cite{martinek:2003,bulka:1999,mishchenko:2003,lamacraft:2004,zareyan:2005} The charge- and the spin-current noise has been considered for zero as well as for finite frequency. There are works that additionally focus on the investigation of the effect of spin-flip scattering on the charge- or spin-current fluctuations.\cite{mishchenko:2003,lamacraft:2004,zareyan:2005} It has been predicted that the Fano factor strongly depends on the lead polarization, the spin-flip scattering strength, and the contact resistances. The limit of arbitrary angle $\phi$ but absence of Coulomb charging effects on the central electrode has been studied in Ref.~\onlinecite{tserkovnyak:2001}.
	
In the present work, we derive a theoretical framework relying on a diagrammatic real-time approach that incorporates the general description of {\it noncollinear} lead magnetization directions and Coulomb charging effects on the central electrode. The used theory allows for a systematic expansion in the tunnel-coupling strength $\Gamma$. Due to the weak coupling between island and leads we perform a perturbation expansion of the transport properties up to first order in $\Gamma$. The presented theory allows for the investigation of the zero-frequency as well as the frequency-dependent current noise for the noncollinear single-electron spin-valve transistor. 

\section{Model}	\label{sec:model}
The system under consideration is the single-electron spin-valve transistor that is illustrated in Fig.~\ref{fig:system}. Its Hamiltonian takes the form
\begin{eqnarray}
	H = \sum_{r=L,R} H_r +H_I + H_C+\sum_{r=L,R} H_{T,r}.
\label{eq:totalH}
\end{eqnarray}

The left ($r=L$) and right ($r=R$) ferromagnetic leads are described as reservoirs of noninteracting electrons by
\begin{equation}\label{eq:hamiltonianLR}
	H_r = \sum_{k{s}\nu} \epsilon_{rk{s}}^{}\, a_{rk{s}\nu}^\dagger a_{rk{s}\nu}\,,
\end{equation}
with $a_{rk{s}\nu}^{(\dagger)}$ being the annihilation (creation) operator of lead $r$, momentum $k$, and transverse-channel index $\nu=1,2,...,N_c$.  The majority (minority) spin states are quantized along the lead magnetization direction ${\bf\hat{n}}_r$ and denoted by the spin index $s=+(-)$. 
The two vectors ${\bf\hat{n}}_L$ and ${\bf\hat{n}}_R$ enclose the angle $\phi$. For simplicity we choose the density of states $\rho_{s}^{r}$, that describes the spin-$\sigma$ electrons in lead $r$, to be energy independent. As a consequence, the lead's degree of spin polarization $p_r= (\rho_{+}^{r}-\rho_{-}^{r})/(\rho_{+}^{r}+\rho_{-}^{r})$ is constant in energy.

\begin{figure}[b]
	\centering
	\includegraphics[width=.8\columnwidth]{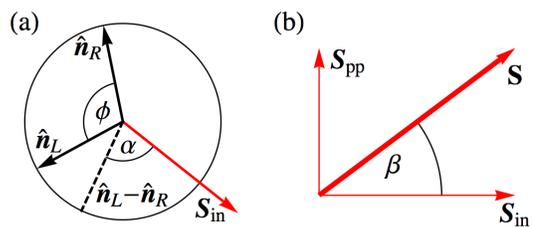}
	\caption{(Color online) Scheme of the relative orientation of the polarization directions of the two leads ${\bf\hat{n}}_L,{\bf\hat{n}}_R$ and the accumulated spin on the island ${\bf S}$, parametrized by the angles $\alpha$ and $\beta$. For clarity reasons, the island spin ${\bf S}$ is decomposed into two parts ${\bf S}={\bf S}_\text{in}+{\bf S}_\text{pp}$, where ${\bf S_\text{in}}$ and ${\bf S}_\text{pp}$ are the contributions in the $({\bf\hat{n}}_L,{\bf\hat{n}}_R)$-plane and perpendicular to it, respectively.}
\label{fig:angles}
\end{figure}

The two contributions $H_I$ and $H_C$ represent the metallic island with an energy spectrum $\vareps_l$ that is characterized by a typical level spacing $\Delta\vareps$. In our work, we are interested in the limit $k_BT,eV\gg \Delta\vareps$, where the spectrum can be viewed as continuous. The first part
\begin{equation}\label{eq:hamiltonianI}
	H_I=\sum_{l\sigma\nu}\vareps_{l}\,c_{l\sigma\nu}^\dagger c_{l\sigma\nu}\,,
\end{equation}
describes the kinetic energy of the electrons with spin $\sigma$ occupying the island level $l$ in the transverse channel $\nu$. The annihilation (creation) operator of island electrons in the state $l\sigma\nu$ is denoted by $c_{l\sigma\nu}^{(\dagger)}$. In our model the levels of the spectrum are assumed to be independent of spin $\sigma$ and channel $\nu$. Due to the noncollinear lead magnetization directions there is no canonical choice for the spin quantization axis of the island. As demonstrated in Ref. \onlinecite{lindebaum:2011} it is convenient to choose the island spin-quantization axis ${\bf\hat{n}}_S$ parallel to the accumulated island spin ${\bf S}$. In Fig.~\ref{fig:angles} we introduce two angles that determine the orientation of ${\bf S}$ relative to the lead magnetization directions.
	The angle $\alpha$ is enclosed by the $({\bf\hat{n}}_L\!\!-\!{\bf\hat{n}}_R)$-axis and the projection of ${\bf S}$ onto the $({\bf\hat{n}}_L,{\bf\hat{n}}_R)$-plane, and $\beta$ is defined as the angle between ${\bf S}$ and the $({\bf\hat{n}}_L,{\bf\hat{n}}_R)$-plane.

The second term describing the island models the Coulomb interaction of electrons that occupy the island:
\begin{equation}\label{eq:hamiltonianC}
	H_C=E_C(N-N_\text{ext})^2\,.
\end{equation}
Here, $N$ is the number of island electrons and the parameter $E_C=e^2/(2C_\Sigma)$ represents the charging-energy scale of the system with $e$ being the elementary charge. The total capacitance $C_\Sigma=C_L+C_R+C_G$ is the sum of the capacitances of the left and right interfaces and the gate. We assume equal junction capacitances ($C_L=C_R$) since they are much less sensitive to the geometry of the tunnel contacts than the tunnel couplings.  Furthermore, the bias voltage $V$ is applied symmetrically to the leads, i.e., their electrochemical potentials are $\mu_L=eV/2$ and $\mu_R=-eV/2$. 
As a consequence, the external charge $e\,N_\text{ext}=C_GV_G$ depends only on the gate  voltage $V_G$ (and not on the bias voltage $V$). For later convenience, we define the energy of the decoupled island system in a microscopic island state $\chi$ as $E_\chi=\bra{\chi}(H_I+H_C)\ket{\chi}$ and additionally we define $\Delta_N$ as the difference of charging energies of $N+1$ and $N$ electrons, i.e., $\Delta_N=E_C [2(N-N_\text{ext})+1]$. 

Electron tunneling between the ferromagnetic lead $r$ and the central island is described by the tunneling Hamiltonian	
\begin{equation}\label{eq:hamiltonianT}
	H_{T,r}=\sum_{kl{s}\sigma\nu}\,V^{r}_{s\sigma}\,a^\dag_{rk{s}\nu} c_{l\sigma\nu} + \text{H.c.}\,,
\end{equation}
where we already assumed that the tunneling-matrix elements $V^{r}_{{s}\sigma}$ are independent of momentum $k$ and transverse channel index $\nu$. Due to the chosen quantization axis of the island spin the elements $V^{r}_{{s}\sigma}$ consist not only of the spin-independent tunnel amplitude $t_r$. In fact they additionally contain matrix elements of an SU(2) rotation that connects the in general different spin quantization axes. In terms of the angles $\alpha$ and $\beta$ the tunneling-matrix elements of the left lead are given by
 \begin{eqnarray}\label{eq:tme}
	V^L_{\pm\up}\!\!\!&=&\!\!\!\frac{t_L}{\sqrt{2}}\left[\pm e^{i\phi /2}\cos\!\!\left(\!\frac{\beta}{2}-\frac{\pi}{4}\!\right)-i e^{i\alpha}\sin\!\!\left(\!\frac{\beta}{2}-\frac{\pi}{4}\!\right)\!\!\right],\\
	V^L_{\pm\dn}\!\!\!&=&\!\!\!\frac{t_L}{\sqrt{2}}\left[\pm e^{i\phi /2}\sin\!\!\left(\!\frac{\beta}{2}-\frac{\pi}{4}\!\right)+i e^{i\alpha}\cos\!\!\left(\!\frac{\beta}{2}-\frac{\pi}{4}\!\right)\!\!\right].
\end{eqnarray}
The elements describing the right lead can directly be obtained by replacing $L\rightarrow R$ and $\phi\rightarrow -\phi$. The tunneling rate for electrons from lead $r$ with spin $s$ into the island spin state $\sigma$ is given by $ \Gamma_{s\sigma}^r/\hbar =  2 \pi \rho_{s}^{r} |V^{r}_{s\sigma}|^2/\hbar $. Finally, we define $\Gamma_\sigma^r=\sum_s\Gamma_{s\sigma}^r$, $\Gamma_r=\sum_\sigma\Gamma_\sigma^r/2$ and $\Gamma=\sum_r \Gamma_r$.

\section{Method}	\label{sec:method}
In Ref.~\onlinecite{lindebaum:2011} we presented a diagrammatic real-time technique to calculate the average current through a noncollinear single-electron spin-valve transistor. The approach incorporates noncollinearity of the lead magnetization directions as well as Coulomb charging effects on the central electrode that is treated nonperturbatively. In order to address the frequency-dependent current fluctuations of the device, we need to extend this theory accordingly, which is described in this section. 
 
The section is divided into three parts covering the derivation of the density matrix that describes the single-electron spin-valve transistor (Sec.~\ref{subsec:denMatr}), of the charge-current (Sec.~\ref{subsec:current}), and of the frequency-dependent current noise (Sec.~\ref{subsec:cuNoise}).

\subsection{Reduced Density Matrix}\label{subsec:denMatr}
Since the leads are considered as equilibrium reservoirs of non-interacting electrons, we integrate them out and derive an effective description, which only contains the degrees of freedom of the metallic island characterized by the reduced density matrix $\hat{\rho}_\text{red}$. For time-translation invariant systems its time evolution can be expressed in terms of the reduced propagator ${\bf \Pi}(t-t_0)$:
\begin{eqnarray}
	\hat{\rho}_\text{red}(t)={\bf \Pi}(t-t_0)\hat{\rho}_\text{red}(t_0).
\end{eqnarray}
The initial density matrix $\hat{\rho}_\text{red}^\text{ini}$ of the system is given by $\hat{\rho}_\text{red}^\text{ini}=\lim_{t_0\rightarrow-\infty}\hat{\rho}_\text{red}(t_0)$. 
In the long-time limit, the system looses any information about its initial state. Hence we are free to define the elements of the initial density matrix as $(\hat{\rho}_\text{red}^\text{ini}){}^{\chi_1}_{\chi_2}=\delta_{\chi_1,\chi_0}\delta_{\chi_2,\chi_0}$ with $\chi_0$ being an arbitrary state of the reduced system. The microscopic island states $\chi$ are determined by $\ket{\chi}=\ket{ \{ n_{l\sigma \nu} \}}$, with $n_{l\sigma \nu}=0,1$ representing whether the corresponding island level $l\nu$ is occupied by a spin-$\sigma$ electron. Furthermore, we define the stationary reduced density-matrix elements as $P^{\chi_1}_{\chi_2}=\bra{\chi_1}\hat{\rho}_\text{red}\ket{\chi_2}$. In the used notation, the diagonal elements $P_\chi^\chi\equiv P_\chi$ correspond to the occupation probabilities of state $\chi$. They fulfill the normalization condition $\sum_\chi P_\chi=1$. Eventually, we get the following equation for the elements of $\hat{\rho}_\text{red}$:
\begin{eqnarray}
	P^{\chi_1}_{\chi_2}=\lim_{t_0\rightarrow-\infty} \Pi(t-t_0)_{\chi_2\, \chi_0}^{\chi_1\,\chi_0}.
\end{eqnarray}
By means of the transform ${\bf \Pi}(\omega)=\int^{\infty}_0\text{d}t{\bf \Pi}(t)\exp[-i(\omega-i0^+)t]/\hbar$ we switch into frequency space which is convenient in the following. Then the Dyson equation ${\bf \Pi}(\omega)={\bf \Pi}^{(0)}(\omega)+{\bf \Pi}^{(0)}(\omega){\bf W}(\omega){\bf \Pi}(\omega)$ yields the following form of the reduced propagator:
\begin{eqnarray}\label{eq:propagator}
	{\bf \Pi}(\omega)=\left({{\bf \Pi}^{(0)}(\omega)}^{-1}-{\bf W}(\omega)\right)^{-1},
\end{eqnarray}
with the free propagator ${\bf \Pi}^{(0)}(\omega)$ and the kernel ${\bf W}(\omega)$. The matrix elements of the former are given by
\begin{eqnarray}	
	\Pi_{\quad\chi_2^{}\chi_2'}^{(0)\chi_1^{}\chi_1'}(\omega)=\frac{i\delta_{\chi^{}_1,\chi'_1}\delta_{\chi^{}_2,\chi'_2}}{E_{\chi_1}-E_{\chi_2}-\hbar\omega+i0^+}.
\end{eqnarray}
They describe free propagation in time while the elements $W_{\chi_2\,\chi'_2}^{\chi_1\,\chi'_1}(\omega)$ characterize transitions between the matrix elements $P_{\chi'_2}^{\chi'_1}$ and $P_{\chi^{}_2}^{\chi^{}_1}$. The kernel ${\bf W}$ can be calculated within a diagrammatic real-time technique that allows for a systematic perturbative expansion in the tunnel-coupling strength $\Gamma$.\cite{koenig:1996,koenig:1996a,schoeller:1997,koenig:1999,braun:2004} In the present work, it is sufficient to expand it up to the first order since we want to describe weak coupling between island and leads. In this situation, sequential-tunneling processes are dominant, however, the formalism is generally formulated and not restricted to the sequential-tunneling limit. We note that in the diagrammatic language, the transformation into frequency space incorporates an additional bosonic line carrying the energy $\hbar\omega$ into the diagrams.

By applying the final value theorem $\lim_{\omega\rightarrow 0}(i\omega+0^+){\bf \Pi}(\omega)=\lim_{t\rightarrow \infty}{\bf \Pi}(t)=\hat{\rho}_\text{red}$ to Eq.~(\ref{eq:propagator}) we, finally, obtain the matrix form of the {\it generalized master equation} that determines $\hat{\rho}_\text{red}$ in the stationary limit
\begin{eqnarray}\label{eq:meFreq}
	0=\left[{\bf \Pi}^{(0)}(\omega=0)^{-1}-{\bf W}(\omega=0)\right]\hat{\rho}_\text{red}.
\end{eqnarray}
Written in component form this equation reads 
\begin{eqnarray}\label{eq:mastereq}
	0=\dt{}{P_{\chi_2}^{\chi_1}}=-\frac{i}{\hbar}\left(E_{\chi_1}-E_{\chi_2}\right)P_{\chi_2}^{\chi_1} +\!\!\!\sum_{\chi'_1\chi'_2}W_{\chi_2\, \chi'_2}^{\chi_1\,\chi'_1}P_{\chi'_2}^{\chi'_1}.
\end{eqnarray}
Although we have already traced out the lead degrees of freedom, due to the continuous island spectrum the system of equations that has to be solved is still highly dimensional. In the rest of this subsection we will follow the procedure of Ref.~\onlinecite{lindebaum:2011} to strongly reduce this large number of degrees of freedom to the relevant ones for electronic transport. The first step is to get rid of the off-diagonal density-matrix elements on the right-hand side Eq.~(\ref{eq:mastereq}). In our system, the only possibility to change the microscopic state of the island is tunneling of electrons from the leads to the central region or vice versa, i.e., we neglect intrinsic spin-flip processes in the island due to a large spin-flip time scale $\tau_\text{sf}$. Tunneling is described by the Hamiltonians $H_{T,r}$ which conserve charge and transverse channel $\nu$. Hence the elements of the reduced density matrix $P_{\chi'}^\chi$ that have to be taken into account fulfill the condition $\sum_{l\sigma} n_{l\sigma\nu}= \sum_{l\sigma} n'_{l\sigma\nu}$. Furthermore, due to large number of relaxation channels in metallic islands we assume that the electron dwell time $\tau_\text{dw}$ is larger than the energy-relaxation time $\tau_\text{er}$. This results in the fact that coherent superpositions between island states that differ in the number of electrons occupying the level $l$ decay quickly such that $\sum_{\sigma} n_{l\sigma\nu}= \sum_{\sigma} n'_{l\sigma\nu}$. With that we already achieved that the first contribution of the right-hand side of Eq.~(\ref{eq:mastereq}) vanishes since spin degeneracy of the island spectrum yields $E_\chi-E_{\chi'}=0$ for all relevant states $\chi$ and $\chi'$. To get rid of all remaining off-diagonal elements of $\hat{\rho}_\text{red}$ in the second contribution of Eq.~(\ref{eq:mastereq}) we neglect quantum corrections to spin quadrupole and higher moments, i.e, only spin-dipole moments of the island are taken into account. This yields $n_{l\sigma\nu}=  n'_{l\sigma\nu}$, and Eq.~(\ref{eq:mastereq}) simplifies to the kinetic equation 
\begin{equation}\label{eq:simpleME}
	0=  \dt{}{P_{\chi_2}^{\chi_1}}= \sum_{\chi} W_{\chi_2 \, \chi}^{\chi_1 \, \chi} P_{\chi}\;.
\end{equation}
	
To describe the electronic structure of the central metallic island it is useful to introduce the spin-dependent electrochemical potentials of the island spins $\mu_\sigma$. These quantities describe the two spin subsystems that are in general out of equilibrium, i.e., $\mu_\up\neq\mu_\dn$. In the considered limit $\tau_\text{sf}\gg\tau_\text{er},\tau_\text{dw}$, the two different spin species act like two independent reservoirs of electrons, which both may be described by the Fermi distribution $f(E-\mu_\sigma)=[\exp(\frac{E-\mu_\sigma}{k_BT})+1]^{-1}$. In general, the spin-dependent chemical potentials $\mu_\sigma$ depend on the number $N_\sigma$ of spin-$\sigma$ electrons occupying the island ($\mu_\sigma=\mu_\sigma(N_\sigma)$) and have to be determined by the expression $N_\sigma = \sum_l f\left[\epsilon_l-\mu_\sigma(N_\sigma)\right]$. But in the case of small level splittings $\Delta\vareps\ll k_BT,eV$ it is reasonable to assume $\mu_\sigma$ to be independent of $N_\sigma$.

The simplified master equation Eq.~(\ref{eq:simpleME}) still depends on the microscopic island states $\chi$ that contain all information about the individual occupation of each island level. However, most of this information is irrelevant for the evaluation of the spin dynamics, the average current, and the current-current correlation function of the single-electron spin-valve transistor. The independent degrees of freedom that have to be taken into account are the three components of the accumulated island spin ${\bf S}$ represented by its magnitude $S=\hbar\rho_I(\mu_\up-\mu_\dn)/2$ and its spacially orientation that is characterized by the angles $\alpha$ and $\beta$ as well as the probabilities to find $N$ electrons on the island. The latter are defined as $P_N = \sum_\chi P_\chi \delta_{N,N_\chi}$ with $N_\chi$ being the number of island electrons in state $\chi$. The equations that enable a calculation of each relevant independent degree of freedom are provided by the kinetic equations of the charging-state projector $\ket{N}\bra{N} = \sum_\chi |\chi\rangle \langle \chi | \delta_{N,N_\chi}$ and the total island spin operator ${\bf \hat{S}} = (\hbar/2) \sum_{l\sigma \sigma' \nu} c_{l\sigma\nu}^\dagger {\vec{\sigma}}_{\sigma \, \sigma'} c_{l\sigma'\nu}$, where $\vec{\sigma}$ is the Pauli spin-matrices vector. One obtains
\begin{eqnarray}
	0=\dt{}P_N &=&\sum_{\chi \chi' } \delta_{N,N_\chi} W_{\chi \, \chi'}^{\chi \, \chi'} P_{\chi'}\;,\\
	0=\dt{}{\langle {\bf \hat{S}} \rangle}&=&\sum_{\chi_1\chi_2\chi'} \bra{\chi_2} {\bf \hat{S}} \ket{\chi_1} W_{\chi_2 \, \chi'}^{\chi_1 \, \chi'} P_{\chi'} \; .
\end{eqnarray}
The elements of the kernel ${\bf W}$ are evaluated with the help of the diagrammatic rules explained in detail in Ref.~\onlinecite{lindebaum:2011}. By inserting the obtained kernel elements and by additionally utilizing the previously discussed model simplifications we eventually obtain the kinetic equations of the independent degrees of freedom $P_N$ and ${\bf S}$. The charging-state occupation probabilities are given by 
\begin{eqnarray}\label{eq:kineticEqPN}	
	\nonumber\dt{}P_{N}&=&\pi\sum_{r\sigma}\left[\alpha_{r\sigma}^{+}(\Delta_{N-1})P_{N-1}+\alpha_{r\sigma}^{-}(\Delta_{N} )P_{N+1}\right.\\
	&&\qquad\left.-\alpha_{r\sigma}^{+}(\Delta_{N} )P_{N}-\alpha_{r\sigma}^{-}(\Delta_{N-1} )P_{N}\right],
\end{eqnarray}
while the time evolution of the accumulated island spin can be expressed as a Bloch-like equation
\begin{eqnarray}\label{eq:kineticEqS}
	\dt{\langle{\bf S}\rangle}=\left(\dt{\langle{\bf S}\rangle}\right)_\text{acc}+\left(\dt{\langle{\bf S}\rangle}\right)_\text{rel}+\left(\dt{\langle{\bf S}\rangle}\right)_\text{rot}.
\end{eqnarray}
The three contributions describe accumulation, relaxation, and rotation of the island spin. Their explicit forms are given in Appendix~\ref{ap:islandSpinKE}. We note that the interaction-induced exchange field that exists between the ferromagnetic leads and the metallic island is contained in the spin-rotation term .

In both sets of the kinetic equations, Eqs.~(\ref{eq:kineticEqPN}) and (\ref{eq:kineticEqS}), the island rate functions
\begin{eqnarray}
	\alpha_{r\sigma}^{\pm}(E):=\pm\alpha_{r\sigma}^0\;\frac{E-(\mu_r-\mu_\sigma)}{\exp\left[\pm\frac{E-(\mu_r-\mu_\sigma)}{k_B T}\right]-1},
\end{eqnarray}
with the dimensionless conductance $\alpha_{r\sigma}^0=\frac{\rho_IN_c}{2\pi\hbar}\Gamma_\sigma^r$ appear. They describe tunneling of spin-$\sigma$ electrons with energy $E$ between lead $r$ and central island. 
 
\subsection{Charge Current}\label{subsec:current}
The charge current operator $I_r=-e \text{d} N_r/ \text{d}t$ through lead $r$, where $N_r$ is the total number of electrons in lead $r$,  is obtained from the quantum-mechanical equation of motion in Heisenberg picture, 
\begin{equation}\label{eq:current}
	I_r=-\frac{ie}{\hbar}[H,N_r]=-\frac{ie}{\hbar}\sum_{kl{s}\sigma\nu}\,V^{r}_{s\sigma}\,a^\dag_{rk{s}\nu} c^{}_{l\sigma\nu} + \text{H.c.}\;.
\end{equation}
A comparison of this expression with the tunneling Hamiltonian of lead $r$, see Eq.~(\ref{eq:hamiltonianT}), yields that $I_r$ can directly be obtained from $H_{T,r}$ by performing the replacement $V^{r}_{s\sigma}\rightarrow -ieV^{r}_{s\sigma}/\hbar$. The constant complex prefactor can easily be incorporated in the used diagrammatic technique. Eventually, the stationary charge current through lead $r$ is given by
\begin{equation}\label{eq:current}
	I_r=\sum_{\chi\,\chi'_1\chi'_2}\left(W^{I_r}\right)_{\chi\, \chi'_2}^{\chi\, \chi'_1}P_{\chi'_2}^{\chi'_1} \, ,
\end{equation}
where the matrix elements of the current transition rates ${\bf W}^{I_r}$ are directly obtained by multiplying the corresponding matrix elements of ${\bf W}$ with the net transported charge from lead $r$ to the island. We note that minus signs originating from the complex prefactor and factors $1/2$ that appear due to the definition of the symmetrized current have to be taken into account appropriatly. 

Since for the average current charge conservation yields $I=I_L=-I_R$, we can consider the symmetrized current \mbox{$I=(I_L-I_R)/2$}, which will be particularly useful for the discussion of the finite-frequency noise, see subsection~\ref{subsec:cuNoise}. The symmetrized current can be written in matrix notation as
\begin{eqnarray}
	I=\text{tr}\left[{\bf W}^I(\omega=0)\hat{\rho}_\text{red}\right].
\end{eqnarray}
By adapting the same procedure that we used in deriving the kinetic equation Eq.~(\ref{eq:simpleME}) to the expression of the current $I$ one gets
\begin{eqnarray}
	I=\frac{e\pi}{2}\sum_{rN\sigma}\pm\left[\alpha_{r\sigma}^{+}(\Delta_N)-\alpha_{r\sigma}^{-}(\Delta_{N-1})\right]P_{N} \, .
\end{eqnarray}
The upper (lower) sign has to be taken for the left (right) lead.

\subsection{Current Noise}\label{subsec:cuNoise}
Fluctuations of the current are described by the current-current correlation function
\begin{eqnarray}
	S^{II}(t)=\mw{I(t)I(0)}+\mw{I(0)I(t)}-2\mw{I}^2.
\end{eqnarray}
Its Fourier transform defines the frequency-dependent current noise:
\begin{eqnarray}
	\!\!\!\!\!\!\!\!\!\nonumber S^{II}(\omega)\!\!\!&=&\!\!\!\!\!\int\limits_{-\infty}^\infty\text{d}t\;S^{II}(t)e^{-i\omega t}\\
	&=&\!\!\!\!\!\int\limits_{-\infty}^\infty\!\!\!\text{d}t\!\left[\mw{I(t)I(0)}+\mw{I(0)I(t)}\right]e^{-i\omega t}\!\!-\!4\pi\!\mw{I}^2\!\delta(\omega).
	\label{eq:freqNoise}
\end{eqnarray}
By definition, $S^{II}(\omega)$ is symmetric in frequency, $S^{II}(\omega)=S^{II}(-\omega)$. It represents a real physical observable. For finite frequencies, the total current flowing through the single-electron spin-valve transistor is not equal to the symmetrized current $I=(I_L-I_R)/2$ that occurs in Eq.~(\ref{eq:freqNoise}). In general, displacement currents appear that have to be taken into account by defining the current as $I=(C_LI_L-C_RI_R)/(C_L+C_R)$.\cite{korotkov:1994} However, in our model we assume symmetric junction capacitances, see section \ref{sec:model}, and hence the general formula corresponds to the definition of the symmetrized current.
	
In the following, we demonstrate how the calculation of the noise $S^{II}(\omega)$ can be realized within the diagrammatic technique that was used to derive the kinetic equations and the charge current formula. In Ref.~\onlinecite{braun:2006}, Braun {\it et al.} developed a formalism relying on the same diagrammatic theory that enables to calculate the frequency-dependent current noise of a quantum-dot spin valve, which is a single-level quantum dot coupled to two ferromagnetic leads. However, due to the large number of levels contributing to transport this theoretical framework is not applicable to our system. Hence we present how we extend the theory to describe the current-current fluctuations of the single-electron spin-valve transistor.

The frequency $\omega$ appearing in Eq.~(\ref{eq:freqNoise}) is taken into account in the diagrammatic language by introducing an additional bosonic line carrying the energy $\hbar\omega$ that connects the two current vertices that replace two vertices originating from the tunneling Hamiltonian.\cite{braun:2006}
 
In matrix notation the frequency-dependent current noise is given by
\begin{eqnarray}
	\nonumber \!\!\!\!\!\!\!\!\!\!\!\!S^{II}(\omega)\!\!\!&=&\!\!\!-4\pi\mw{I}^2\delta(\omega)\\
	\nonumber&&\!\!\!+\frac{1}{2}\sum_{\gamma=\pm}\tr{{\bf W}^{II}(\gamma\omega)\hat{\rho}_\text{red}\right.\\
	&&\qquad\qquad\;\left.+{\bf W}^I_<(\gamma\omega){\bf \Pi}(\gamma\omega){\bf W}^I_>(\gamma\omega)\hat{\rho}_\text{red}}.
\label{eq:generalS}
\end{eqnarray}
with three new diagrammatic objects ${\bf W}^{II}(\omega)$, ${\bf W}^I_<(\omega)$, and ${\bf W}^I_>(\omega)$.
 
In ${\bf W}^I_<(\omega)$ and ${\bf W}^I_>(\omega)$ one current vertex placed on the upper or lower Keldysh contour is contacted by the bosonic line that enters the diagrams from the left or leaves them to the right, respectively. In the diagrams of ${\bf W}^{II}(\omega)$ the additional line connects two current vertices.

The noise formula given in Eq.~(\ref{eq:generalS}) represents the general expression of the current fluctuations of the single-electron spin-valve transistor. 
In the limit of weak island-lead coupling, only diagrams containing one tunneling line contribute to the kernels, i.e., they are expanded up to first order in $\Gamma$. 
For a systematic and consistent perturbation expansion of the expression for the noise, we count the frequency as one order in the tunnel-coupling strength, $\omega\sim\Gamma$, and expand to first order in this small parameter. In this limit the frequency dependence of the kernels can be neglected since each correction in $\omega$ evokes contributions that are at least proportional to $\Gamma^2$. The only frequency dependence we keep is contained in the reduced propagator ${\bf \Pi}(\omega)$, see Eq.~(\ref{eq:propagator}). This expression is treated consistently when only the frequency dependence of the free propagator is taken into account. As a result, we find
\begin{eqnarray}
	\nonumber \!\!\!\!\!\!\!\!\!\!\!\!S^{II}(\omega)\!\!\!&=&\!\!\!-4\pi\mw{I}^2\delta(\omega)\\
	\nonumber&&\!\!\!+\frac{1}{2}\sum_{\gamma=\pm}\!\tr{{\bf W}^{II}\hat{\rho}_\text{red}\right.\\
	&&\qquad\qquad\left.+{\bf W}^I\!({{\bf \Pi}^{(0)}(\gamma\omega)}^{-1}\!\!\!-{\bf W})^{-1}{\bf W}^I\hat{\rho}_\text{red}}\!,
\label{eq:generalSGamma}
\end{eqnarray}
with the definitions ${\bf W}={\bf W}(\omega=0)$, ${\bf W}^I={\bf W}^I_<(\omega=0)={\bf W}^I_>(\omega=0)$, and ${\bf W}^{II}={\bf W}^{II}(\omega=0)$. This formula for $S^{II}(\omega)$ depends on all the elements of the reduced density matrix $P^{\chi_1}_{\chi_2}$. 
An effective description that only contains the charge-state occupation probabilities $P_N$ and the accumulated island spin ${\bf S}$ as degrees of freedom is obtained by executing the same procedure as in the derivation of the kinetic equations. This enables to remove the coherent superpositions on the right-hand side of Eq.~(\ref{eq:generalSGamma}), i.e., only diagonal matrix elements $P_\chi$ enter. In the island-charge-state basis the frequency dependent noise is then given by
\begin{eqnarray}
	\nonumber \!\!\!\!\!\!\!\!\!\!\!S^{II}(\omega)\!\!\!&=&\!\!\!-4\pi\mw{I}^2\delta(\omega)+{\bf e}^\text{T}\widetilde{{\bf W}}^{II}{\bf P}\\
	&&\!\!\!+\frac{1}{2}{\bf e}^\text{T}\!\!\left[\sum_{\gamma=\pm}\widetilde{{\bf W}}^I({\widetilde{{\bf \Pi}}^{(0)}(\gamma\omega)}^{-1}\!\!\!-\!\widetilde{{\bf W}})^{-1}\widetilde{{\bf W}}^I\right]\!\!{\bf P},
\label{eq:generalSN}
\end{eqnarray}
where the vector ${\bf e}$ is defined by $e_N=1$ for all $N$ and the vector of the island-occupation probabilities ${\bf P}=(..,P_{N-1},P_{N},P_{N+1},...)$ fulfills the normalization condition ${\bf e}^\text{T}{\bf P}=1$. The matrix elements of the kernels in charge space that have to be plugged in the formula Eq.(\ref{eq:generalSN}) are defined as follows:
\begin{eqnarray}
	\nonumber\widetilde{W}_{N,N'}\!\!\!&=&\!\!\!\pi\sum_{r\sigma}\left\{\delta_{N,N'+1}\alpha^+_{r\sigma}(\Delta_{N-1})+\delta_{N+1,N'}\alpha^-_{r\sigma}(\Delta_{N})\right.\\
	&&\left.\qquad\;-\delta_{N,N'}\left[\alpha^+_{r\sigma}(\Delta_{N})-\alpha^-_{r\sigma}(\Delta_{N-1})\right]\right\},\label{eq:WTilde}\\
	\nonumber\widetilde{W}^I_{N,N'}\!\!\!&=&\!\!\!e\pi\sum_{r\sigma}\pm\left[\delta_{N,N'+1}\alpha^+_{r\sigma}(\Delta_{N-1})\right.\\
	&&\left.\qquad\quad-\delta_{N+1,N'}\alpha^-_{r\sigma}(\Delta_{N})\right],\label{eq:WITilde}\\ 
	\nonumber\widetilde{W}^{II}_{N,N'}\!\!\!&=&\!\!\!\frac{e^2\pi}{2}\sum_{r\sigma}\left[\delta_{N,N'+1}\alpha^+_{r\sigma}(\Delta_{N-1})\right.\\
	&&\left.\qquad\quad+\delta_{N+1,N'}\alpha^-_{r\sigma}(\Delta_{N})\right].
	\label{eq:WIITilde}
\end{eqnarray}
In Eq.~(\ref{eq:WITilde}) the upper/lower sign has to be chosen for the kernel contributions representing the current through the left/right tunnel junction $(r=L/R)$.

As mentioned above, in the considered limit the only frequency dependence that is contained in the noise is represented by the free propagator of the system. In charge space it is given by
\begin{eqnarray}\label{eq:freePrpAp}
	\widetilde{\Pi}^{(0)\;N_1}_{\quad \,N_2\,N'}(\omega)=\frac{i}{-\hbar\omega+i0^+}\delta_{N_1,N'}\delta_{N_2,N'}.
\end{eqnarray}
Hence $\widetilde{{\bf \Pi}}^{(0)}(\omega)$ is represented by a diagonal matrix. For finite frequencies, the $i0^+$ in Eq.~(\ref{eq:freePrpAp}) drops together with the delta function in Eq.~(\ref{eq:generalSN}).

In the Eqs.~(\ref{eq:WTilde})-(\ref{eq:WIITilde}) one finds that the kernels $\widetilde{{\bf W}}$, $\widetilde{{\bf W}}^I$, and $\widetilde{{\bf W}}^{II}$ are proportional to the island rate functions $\alpha^\pm_{r\sigma}$, which makes the expression of the current noise reliable for frequencies $\omega\lesssim\alpha^\pm_{r\sigma}$.

\section{Results}\label{sec:results}
In the present section the results of our calculations concerning the current fluctuations of a single-electron spin-valve transistor are presented. We separately discuss the zero-frequency limit and the finite-frequency noise of the system in subsections \ref{subsec:zeroFreq} and \ref{subsec:finiteFreq}, respectively. For simplicity we assume that the ferromagnetic leads are symmetrically polarized $p_L=p_R=p$, which can be experimentally realized by using electrodes of the same material. However, the tunnel-coupling strengths to the leads $\Gamma_r$ are not restricted to a symmetric setup and we define the asymmetry parameter $a\equiv\Gamma_L/\Gamma_R$.

\subsection{Zero-Frequency Current Noise}\label{subsec:zeroFreq}
\begin{figure}[tb]
	\centering	
 	\includegraphics[width=.9\columnwidth]{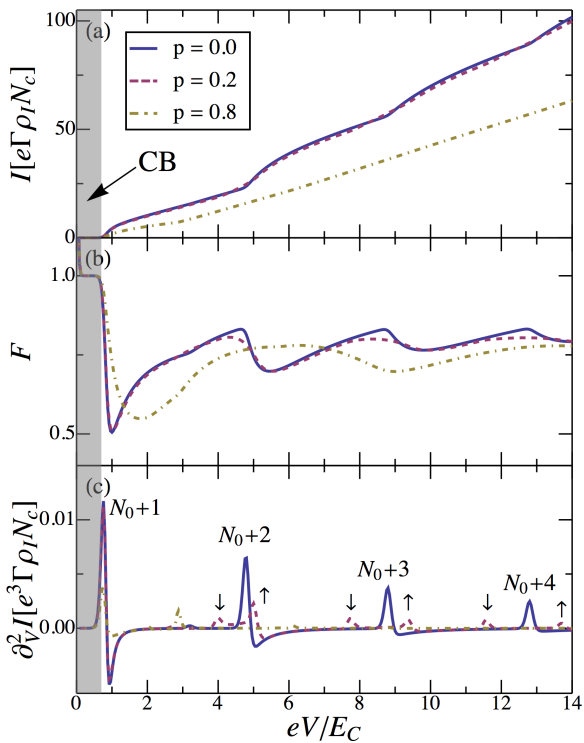}
	\caption{Antiparallel setup ($\phi=\pi$) of the single-electron spin-valve transistor: (a) average current, (b) Fano factor, and (b) second derivative of the current as a function of the applied bias voltage $V$ for different lead polarizations $p$. For all three plots, the remaining parameters were chosen to be $C_GV_G=3e/10$, $E_C=50k_BT$, and $a=10$.}
	\label{fig:res2_ZeroFanoV}
\end{figure}

We start with the case of collinear polarization of the ferromagnets. Due to the symmetric degree of polarization there is no spin accumulation on the central island for parallelly aligned lead magnetization directions ($\phi=0$). This results in the fact that both the current and the noise are independent of the lead polarization and correspond to those of the unpolarized case.\cite{hershfield:1993,bagrets:2003} In the more general setup $p_L\neq p_R$ there is a finite island spin accumulation which results in a polarization dependent Fano factor. The situation changes for the antiparallel setup ($\phi=\pi$) as in this case the spin accumulation strongly depends on $p$. In Fig.~\ref{fig:res2_ZeroFanoV} we show for the antiparallel case (a) the current, (b) the Fano factor, and (c) the second derivative of the current for different values of the lead polarization and a large asymmetry parameter ($a=10$).
We want to emphasize that for obtaining reliable results in the Coulomb-blockade regime, where sequential-tunneling transport is completely blocked (low bias voltages), cotunneling processes have to be taken into account. This regime is marked by the grey area. For vanishing polarization the sub-Poissonian Fano factor exhibits strongly pronounced Coulomb blockade oscilliations which represent the steps in the current-voltage characteristics. Hence the maxima of these oscillations occur at the excitation energies of the relevant island charging states of the central island. The sharp structure of the $p=0$ Fano factor smears out for finite polarizations. Additionally, the position of the maxima change. Both effects are caused by the island spin accumulation since it induces the spin dependence of the charging-state excitation energies that results in a more complex excitation spectrum. To illustrate this behavior we plotted the second derivative of the current in Fig.~\ref{fig:res2_ZeroFanoV}(c) and consider the case of $p=20\%$. The Coulomb peaks of the unpolarized situation are split up into two peaks representing the two different spin reservoirs of the island (marked by the up and down arrows). We emphasize that the first peak ($N_0+1$) does not split due to the exponentially suppressed island spin accumulation in the Coulomb-blockade regime. For the charging states $N_0+n$ with $n\in\mathbb{N}^+$ the spin-down (spin-up) excitation energies are shifted towards lower (higher) bias voltages and vice versa for $N_0-n$. However, due to the asymmetric coupling the occupation of the latter states is strongly suppressed.

Even for very large asymmetry parameters $a\gg1$ and highly polarized materials $p>90\%$ the obtained Fano factors remain sub-Poissonian. This is in contrast to systems consistent of two ferromagnetic leads tunnel coupled to a central region that exhibits a discrete energy spectrum.\cite{bulka:1999,braun:2006,lindebaum:2009} In the case of a single-level quantum dot coupled to two ferromagnetic leads (quantum-dot spin valve) super-Poissonian statistics arises as a result of bunching effects that are caused by spin blockade. In the single-electron spin-valve transistor such bunching effects do not occur since its transport behavior is significantly different. To illustrate this we start by considering the quantum-dot spin valve with parallel aligned lead magnetization directions in the regime where the single level of the dot is predominant occupied by one electron. Finite polarizations evoke that mainly majority spins enter the dot. These can easily leave into the drain lead due to the parallel lead magnetizations. However, if a minority spin occupies the single level then its large dwell time (few minority states in the drain electrode) leads to a temporary blockade of the current through the system since there are no further charging states in the transport window. Hence the rare event of a minority charge carrier tunneling onto the dot bunches the flow of majority electrons. This behavior leads to super-Poissonian noise. In contrast to this, the continuous level structure of the single-electron spin-valve transistor prevents bunching caused by minority electrons. This is due to the fact that there are two different processes that change the charging state of the system after a minority electron tunneled onto the island. On the one hand the same electron or another minority charge carrier can leave the central electrode into the drain with a small probability (similar to the quantum-dot spin valve). But on the other hand, contrary to the single-level system, it is additionally possible that also majority electrons can tunnel into the drain. Even for higher polarizations the latter process is not suppressed. It transfers the system into the lower island charging state and enables subsequent repetitions of island filling and depletion processes. Therefore, the transport through the system is not blocked by minority electrons as in the case of single-level quantum dots, arising in sub-Poissonian Fano factors.
		
Having discussed the limit of parallel and antiparallel magnetization directions we now turn to the noncollinear single-electron spin-valve transistor. If one neglects the Coulomb interaction of electrons on the central island and additionally assumes that the leads are held at zero temperature then the following analytic expression of the Fano factor is obtained:\cite{tserkovnyak:2001}
\begin{eqnarray}
	F=\frac{1}{2}\left(1+p^2\sin^2\frac{\phi}{2}\right).
\label{eq:NoECFano}
\end{eqnarray}
In this limit, the Fano factor exclusively depends on the polarization of the leads and on the angle $\phi$ between the $p$-vectors. The formula represents a monotonically increasing $F$ from its minima $F(0)=1/2$ to its maximal value $F(\pi)=(1+p^2)/2$. However, by neglecting the Coulomb-repulsion energy crucial single-charging effects that govern the transport characteristics of the system are not taken into account. Hence Eq.~(\ref{eq:NoECFano}) is an insufficient description of the noise in a single-electron spin-valve transistor.
		
\begin{figure}[tb]
	\centering
 	\includegraphics[width=.9\columnwidth]{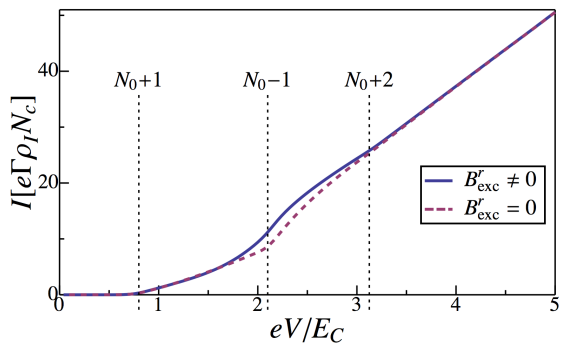}
	\caption{Current through the noncollinear ($\phi=\pi/2$) single-electron spin-valve transistor with (solid) and without (dashed) the effect of the exchange field. The dashed vertical lines represent the threshold voltages of different charging states. The remaining parameters were chosen to be $p=9/10$, $C_GV_G=3e/10$, $E_C=50k_BT$, and $a=1$.}
\label{fig:res2_CurrentV}
\end{figure}

In our formalism the electron-electron interaction on the central island is taken into account nonperturbatively. Its interplay with the finite spin polarization is giving rise to the exchange field that acts on the accumulated spin on the central island. In the following the impact of the field on the current fluctuations of the system is discussed in detail. The exchange field is an immanent part of the used theoretical framework and occurs in the kinetic equation of the island spin, see Eqs.~(\ref{eq:kineticEqS}) and (\ref{eq:meSrot}). The expression of the exchange field contribution that is induced between island and lead $r$ is given by Eq.~(\ref{eq:ExcF}). Due to the lead Fermi functions that appear in the integral the exchange field depends on the applied bias voltage. Therefore, to investigate how the exchange field affects the current noise we have to identify a voltage regime where the field noticeably influences the transport. To this end, we consider the current through the system as a function of the bias voltage for a noncollinear angle $\phi=\pi/2$, see Fig.~\ref{fig:res2_CurrentV}.
In the presented figure the influence of the exchange field is illustrated by the difference of the two shown graphs. Here, the dashed line represents the current of an artificial situation where both exchange-field kinetic-equation contributions ${\bf B}^{r}_\text{exc}$ were manually set to zero. While for the calculation of the solid line the field is fully taken into account. A comparison of the two graphs yields that in the vicinity of the threshold voltage, that enables the occupation of the charging state $N_0-1$, the field strongly affects the current-voltage characteristics. Here, we point out that to observe a significant effect of the exchange field highly polarized leads with $p\gtrsim 0.7$ have to be considered. 
		
\begin{figure}[tb]
 	\centering
	\includegraphics[width=.9\columnwidth]{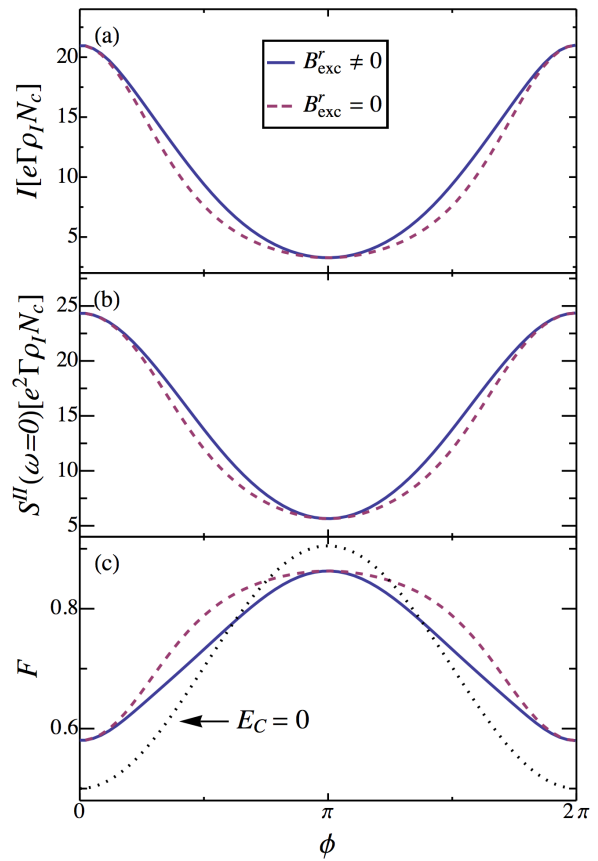}
	\caption{Noncollinear single-electron spin-valve transistor:\newline (a) charge current, (b) zero-frequency noise and (c) Fano factor as a function of the angle between the lead polarization directions $\phi$ with (solid) and without (dashed) the effect of the exchange field. The parameters were chosen to be $p=9/10$,  $eV=2E_C$, $C_GV_G=3e/10$, $E_C=50k_BT$, and $a=1$.}
	\label{fig:res2_NoisePhiFanoPhi}
\end{figure}

After identifying a voltage regime where the exchange field is pronounced we now describe its influence on the current fluctuations of the single-electron spin-valve transistor.  In Fig.~\ref{fig:res2_NoisePhiFanoPhi} the charge current, the zero-frequency noise, and the Fano factor are plotted as a function of the angle between the lead polarization directions $\phi$.
All three plots are calculated for symmetric tunnel-coupling strengths ($a=1$) and a bias voltage of $V=2E_C/e$ that lies in the voltage window exhibiting pronounced exchange fields. The precession of the island spin caused by the field leads to an increase of current and noise. However, the enhancement of the latter is weaker for all angles $\phi$ and therefore the spin precession results in an decreased sub-Poissonian Fano factor. This behavior is also present in systems with asymmetry parameters $a$ being unequal to one. The plotted Fano factors exhibit a monotonic behavior between the two collinear situations. We emphasize that this is caused by the special choice of the parameters, i.e., there are parameter sets for which the maxima (minima) of $F$ do not occur at the collinear angles. In Fig.~\ref{fig:res2_NoisePhiFanoPhi}(c), we additionally plotted the graph of the Fano factor of the noninteracting situation ($E_C=0$) given by Eq.~(\ref{eq:NoECFano}). The comparison shows that a consideration of the noninteracting limit is insufficient to obtain reliable results of the current fluctuations of the single-electron spin-valve transistor. Even for collinear setups the interacting Fano factor strongly deviates from the noninteracting one.

It is worth to mention, that in the exact parallel and antiparallel situations the exchange field contributions of both ferromagnetic leads point in the same direction as the spin accumulation and hence the spin does not rotate in ${\bf B}^{L}_\text{exc}+{\bf B}^{R}_\text{exc}$, i.e., the respective graphs coincide. 
		
\subsection{Finite-Frequency Noise}\label{subsec:finiteFreq}
\begin{figure}[htb]
 	\includegraphics[width=.95\columnwidth]{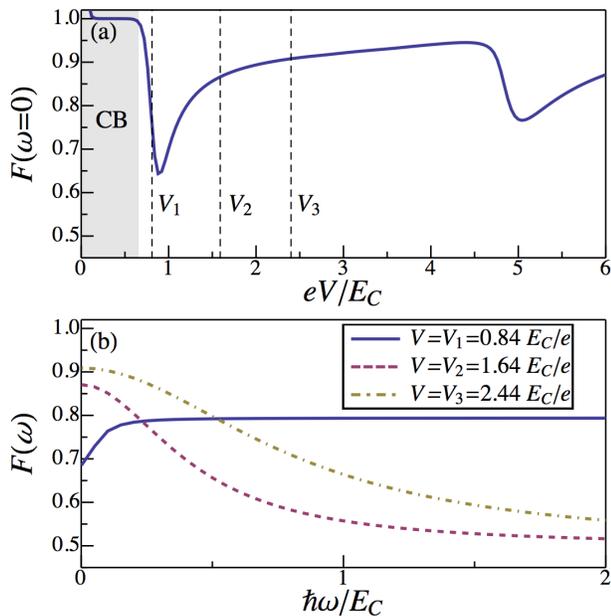}
	\caption{Current fluctuations of the noncollinear ($\phi=\pi/2$) single-electron spin-valve transistor: (a) Zero-frequency Fano factor as a function of bias voltage $V$. (b) Frequency-dependent Fano factor as a function of $\omega$ for different $V$. For both plots the parameters were chosen to be $p=2/10$, $C_GV_G=3e/10$, $E_C=50k_BT$, and $a=40$.}
\label{fig:res2_zeroFFDF}
\end{figure}

To analyze the frequency dependence of the current noise we consider $F(\omega)$ for different bias voltages. The result is shown in Fig.~\ref{fig:res2_zeroFFDF}.
We restrict our investigations to a bias voltage range where the island is exclusively occupied by $N_0$ or by $N_0+1$ electrons. In the upper plot the zero-frequency Fano factor is shown and the different voltages $V_i$ with $i=1,2,3$ for which we will study the current noise are marked by the dashed black lines in Fig.~\ref{fig:res2_zeroFFDF}(a). The energy $eV_1=0.84E_C$ is slightly larger than the excitation energy of the charging state $N_0+1$, i.e., it represents a Coulomb step in the current-voltage characteristics. 
The other two voltages are lying in between two charging steps and the condition $P_{N_0}+P_{N_0+1}=1$ is fulfilled. The respective frequency-dependent Fano factors (fixed $V_i$) are plotted as a function of the frequency $\omega$ in Fig.~\ref{fig:res2_zeroFFDF}(b). For all voltages the Fano factors are composed of a dynamical frequency-dependent contribution reflecting correlations ($V_2,V_3$) and anticorrelations ($V_1$) as well as a constant contribution. The explicit value of the latter is given by $S^{II}(\omega\gg E_C/\hbar)$. A comparison shows that for larger voltages the noise spectra are broadened and the constant term is shifted below the value of the zero-frequency noise to current ratio.
		 
\begin{figure}[tb]
 	\includegraphics[width=.95\columnwidth]{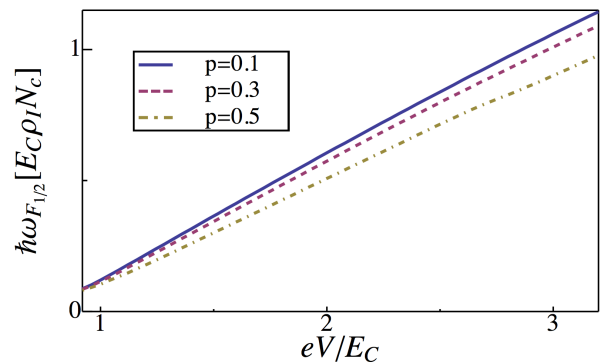}
	\caption{Half width of the frequency-dependent Fano factor as a function of $V$ for different polarizations $p$. The frequency is measured in units $E_C\rho_IN_c$ and the remaining parameters were chosen to be $\phi=\pi/2$, $C_GV_G=3e/10$, $E_C=50k_BT$, and $a=40$.}
\label{fig:res2_FHalfWidth}
\end{figure}

All three graphs in Fig.~\ref{fig:res2_zeroFFDF}(b) indicate that the structure in the noise is destroyed for large enough frequencies. Hence we consider the half width $\omega_{F_{1/2}}$ of the frequency dependent Fano factor as a function of the applied bias voltage to determine the scale on which the system looses its correlation information, see Fig.~\ref{fig:res2_FHalfWidth}.
In the considered voltage regime where only two charging states are allowed the half width is linear in $V$. The slope of the obtained straight line is proportional to $\pi \alpha_0$, with the used definition $\alpha_0\equiv\sum_{r\sigma}\alpha_{r\sigma}^0$. It is maximal for $p=0$ and deacreases with increasing degree of lead polarization. Here, we emphasize that as soon as the voltage is large enough to bring additional higher charging states into the transport window this simple behavior of $\omega_{F_{1/2}}$ is not valid anymore.

\section{Conclusion}    \label{sec:conclusion}
We have presented a diagrammatic real-time technique to investigate the current fluctuations of the single-electron spin-valve transistor in the limit of weak tunnel coupling. The theory allows for {\it noncollinear } lead magnetization directions and simultaneously takes Coulomb charging effects on the central electrode into account.

For the collinear as well as for the noncollinear setup sub-Poissonian transport statistics were observed. We found that finite lead polarizations smear out the characteristic Coulomb blockade oscillations of the Fano factor as a function of the bias voltage. The origin of this behavior was addressed to the splitting of the charging-state excitation energies caused by the nonvanishing spin accumulation on the metallic island. Furthermore, we identified the voltage regime in which the effect of the exchange field is pronounced and demonstrated that ${\bf B}_\text{exc}^r$ leads to a reduction of the zero-frequency Fano factor.

Finally, we analyzed the frequency dependence of the current noise and found that $F(\omega)$ reflects correlations and anticorrelations. However, this noise structure is lost for high enough frequencies. We identified the frequency scale for the corresponding crossover.

\acknowledgments
We acknowledge financial support from DFG via grant KO 1987/4. 

\appendix

\section{Kinetic equations of island spin} \label{ap:islandSpinKE}
The three different contributions of the kinetic equation for the island spin describing accumulation, relaxation, and rotation processes are found to be
\begin{eqnarray}
	\nonumber\left(\dt{\langle{\bf S}\rangle}\right)_\text{acc}\!\!\!\!&=&\!\!\!\frac{\pi\hbar}{2}\sum_{Nr\sigma}p_r\frac{\Gamma_r}{\Gamma_\sigma^r}\left[\hat{{\bf n}}_r+(\hat{{\bf n}}_r\cdot\hat{{\bf n}}_S)\hat{{\bf n}}_S\right]\\
	&&\times\left[\alpha_{r\sigma}^{-}(\Delta_{N-1} ) -\alpha_{r\sigma}^{+}(\Delta_{N} )\right] P_N,\\
	\label{eq:meSacc}
	\nonumber\\
	\nonumber\left(\dt{\langle{\bf S}\rangle}\right)_\text{rel}\!\!\!\!&=&\!\!\!-\!\sum_{Nr\nu}P_N\!\!\int\!\!\text{d}\omega\;\Gamma_r{\bf s}(\omega)\\
	&&\!\!\!\times\left[f_{r}^{-}(\omega+\Delta_{N-1} )-f_{r}^{+}(\omega+\Delta_N )\right]\!,\\
	\label{eq:meSrot}
	\nonumber\\
	\left(\dt{\langle{\bf S}\rangle}\right)_\text{rot}\!\!\!\!&=&\!\!-\frac{g\mu_B}{\hbar}\sum_{r}\int\!\!\text{d}\omega\;{\bf s}(\omega)\times{\bf B}^{r}_\text{exc}(\omega ),
\end{eqnarray}
with the Bohr magneton $\mu_B$, the dimensionless magnetic moment of electrons $g$, and the notation $f_r^+(E)=f(E-\mu_r)$ or $f_r^-(E)=1-f(E-\mu_r)$. Furthermore, the definitions of the energy-dependent spin density in the island ${\bf s}(\omega)=\frac{\hbar\rho_I}{2}[f(\omega-\mu_\up)-f(\omega-\mu_\dn)] \hat{{\bf n}}_S$ and the interaction induced exchange field between central electrode and lead~$r$
\begin{eqnarray}\label{eq:ExcF}
	\!\!\!\!\!\!\!\!\!{\bf B}^{r}_\text{exc}(\omega)\!\!\!&=&\!\!\!\frac{p_r\Gamma_r N_c}{2\pi g\mu_B}\hat{{\bf n}}_r\sum_{N}P_{N}\nonumber \\
	&&\!\!\!\times\!\!\!\int' \!\!\!\text{d}\omega'\!\!\left[\frac{f_r^-(\omega')}{\omega'-\omega-\Delta_N}+\frac{f_r^+(\omega')}{\omega'-\omega-\Delta_{N-1} }\right]\!,
\end{eqnarray}
have been used. A detailed discussion of the different spin contributions and the exchange field is presented in Ref.~\onlinecite{lindebaum:2011}.

\end{document}